\documentclass[accepted=2026-01-26,a4paper,twocolumn,11pt]{quantumarticle}
\pdfoutput=1
\usepackage[utf8]{inputenc}
\usepackage[english]{babel}
\usepackage[T1]{fontenc}
\usepackage{amsmath}
\usepackage{amssymb}
\usepackage{hyperref}

\usepackage{tikz}
\usepackage{lipsum}
\usepackage{booktabs}
\usepackage{xcolor} 

\newcommand{\doiurl}[1]{\href{https://doi.org/#1}{\nolinkurl{https://doi.org/#1}}}

\newcommand{\rev}[1]{#1}

\begin{document}

\title{High-efficiency vertical emission spin-photon interface for scalable quantum memories}

\author{Siavash Mirzaei-Ghormish}
\email{smirzaei@byu.edu}
\affiliation{Department of Electrical and Computer Engineering, Brigham Young University, Provo, UT}

\author{Jeddy Bennett}
\affiliation{Department of Mathematics, Brigham Young University, Provo, UT}

\author{Ryan M. Camacho}
\email{camacho@byu.edu}
\affiliation{Department of Electrical and Computer Engineering, Brigham Young University, Provo, UT}

\maketitle

\begin{abstract} 
We present an efficient spin-photon interface for free-space vertical emission coupling. Using a \rev{dipole model}, we show that our design achieves a far-field collection efficiency of 96\%  at the numerical aperture of 0.7 with a 95\% overlap to a Gaussian mode. 
Our approach is based on a dual perturbation layer design. The first perturbation layer extracts and redirects the resonant mode of a diamond microdisk resonator around the optical axis. The second perturbation layer suppresses side lobes and concentrates most of the light intensity near the center. This dual-layer design enhances control over the farfield pattern and also reduces alignment sensitivity. Additionally, the implemented \rev{dipole model} performs calculations \( 3.2 \times 10^6 \) times faster than full-wave FDTD simulations. These features make the design promising for quantum information applications.
\end{abstract}

\section{Introduction}

Quantum information systems hold tremendous promise for a range of applications that rely on the transmission of quantum states across long distances \cite{kimble2008quantum,gottesman2012longer, bersin2024telecom}. A critical component of such systems are quantum memories with efficient, high-fidelity photon interfaces.  Photonic interfaces to spin-based quantum memories have been demonstrated to be stable, reliable, and scalable \cite{chen2023zero, li2024heterogeneous}.  Solid-state systems, especially color centers in diamonds, have emerged as promising candidates as stationary qubits for interfering with flying qubits.
However, a significant limitation of these color centers, including Nitrogen-Vacancy (NV), Silicon-Vacancy (SiV), and Tin Vacancy (SnV) centers, is their suboptimal emission efficiency into preferred spatial and spectral modes, especially at the zero-phonon line (ZPL) frequency. For example, the ZPL efficiency for NV is 0.03 \cite{iwasaki2017tin}, for SiV is 0.7, and for SnV is 0.8 \cite{gorlitz2020spectroscopic}.  

A potential solution is to embed these color centers into a cavity to enhance their ZPL emission through strong light-matter interactions. Various cavities have been used to enhance the performance of quantum emitter spin-photon interfaces, including one- and two-dimensional photonic crystal cavities \cite{burek2014high, wan2018two, mouradian2017rectangular, saggio2024cavity}, ring resonators \cite{hausmann2014diamond}, fiber-based Fabry--Pérot cavities \cite{hausmann2014diamond,ruf2021resonant, janitz2015fabry}, and circular grating outcouplers \cite{li2015efficient}.
Previous work has independently demonstrated high spectral efficiency and efficient free-space coupling. However, a significant challenge remains: achieving both high spectral and spatial efficiencies simultaneously, ensuring that the majority of photons emitted within the ZPL spectrum are effectively collected and utilized by external optics.

To achieve optimal performance for a spin-photon interface, two key cavity parameters need to be optimized: spectral efficiency  ($\eta_{ZPL}$) and spatial efficiency ($\eta_{col}$) which together determine the figure of merit $\eta_{tot} = \eta_{ZPL} \eta_{col}$.

The spectral efficiency of a quantum emitter within a cavity can be enhanced by Purcell factor \(F_p\) as defined in \cite{li2015coherent}:

\begin{equation}
    \eta_{ZPL} = \frac{F_P} {F_p + \left(\frac{\Gamma_{tot,0}}{\Gamma_{ZPL,0}} - 1 \right)},
    \label{equation1}
\end{equation}

where $\Gamma_{ZPL,0}$ and $\Gamma_{tot,0}$, respectively are the emission rate into the ZPL and the total emission rate without cavity enhancement.  In the case of perfect cavity-atom coupling  $F_p = (3/4\pi^2) (\lambda/n_{eff})^3 (Q/V)$  is the Purcell factor of the ZPL emission, which depends on the cavity's quality factor $Q$ (temporal mode confinement), and the mode volume $V$ (spatial mode confinement).

Once a ZPL photon is emitted into the cavity, it must then be extracted from the cavity and collected them via photonic integrated circuits or free-space optics—becomes.  As indicated by equation \ref{equation1}, optimizing ZPL efficiency requires increasing the Purcell factor. This enhancement can be achieved by designing a cavity with a high Q-factor and a small mode volume. However, there is a trade-off: cavity modes with high Purcell factors are typically highly confined within the cavity. This strong confinement significantly reduces the ability to extract photons efficiently. 

In the saturation regime \(F_p \gg (\frac{\Gamma_{tot,0}}{\Gamma_{ZPL,0}} - 1)\), the difference in ZPL efficiencies between moderate (10-100) and high ($>$100) Purcell factors is minimal. Notably, the efficiency of SnV exceeds 99\% even at relatively low Purcell factors. Whispering gallery mode (WGM) cavities are well-suited for achieving this performance. These cavities show high Q-factors ($10^4$) at the expense of relatively large mode volumes (\( >3(\lambda/n_{eff})^3\)) which leads to moderate Purcell factor. 

To enhance collection efficiency and achieve directional out-coupling, microdisk resonators can be designed with refractive index perturbations, introduced by etching holes into the surface \cite{duan2021vertically}.
This technique optimizes the far-field emission pattern by carefully tuning the perturbation. However, while this method increases collection efficiency, it typically reduces the cavity's quality factor.  Additionally, the process of etching diamond presents significant practical challenges. An enhanced design is proposed in \cite{flores2024alignment}, where, rather than etching directly onto the surface of the diamond cavity, perturbations are introduced by adding an extra layer in the near field above the cavity. While this modification reduces the complexity of fabrication and is  robust for manufacturing variances, the overall collection efficiency in these designs remains limited to approximately 40-50\%.

A Gaussian mode profile is preferred in the far-field, as it aligns well with the fundamental mode of optical fibers. However, many previous designs lack this critical feature \cite{wan2018two,mouradian2017rectangular,li2015efficient,duan2021vertically,flores2024alignment}. Additionally, most previous designs require precise positioning of the quantum emitter within the cavity to achieve high efficiencies \cite{wan2018two,mouradian2017rectangular,li2015efficient,duan2021vertically,flores2024alignment, claudon2010highly, reimer2012bright, gschrey2015highly}. However, the orientation of color centers is often arbitrary, making it difficult to control.

In this paper, we present a design for the vertical emission of spin-polarization entangled photons from SnV centers. The design achieves a high collection efficiency of 96\% within a numerical aperture (NA) of 0.7, with a far-field profile that is 95\% matched to a Gaussian profile. In addition, this design is able to transfer the information to the far-field even when the quantum emitter’s orientation within the cavity is arbitrary. 

To achieve high efficiency and the desired far-field pattern, we employ two perturbation layers: one in the near field and the other in the intermediate field. The first perturbation layer reshapes the near-field into a concentric profile at low NA, while the second acts as a focusing lens, concentrating most of the light around the optical axis. Additionally, the second layer functions as near-lossless a spatial filter, suppressing intensity at high NAs by destructive interference. This approach provides more degrees of freedom to achieve the desired field pattern. Incorporating these two perturbation layers also helps minimize sensitivity to alignment variations between the microdisk resonator and the grating layers.

To assist with the design, we extend the capabilities of our previously published quantum \rev{dipole model} \cite{flores2024alignment}, which calculates and optimizes both collection efficiency and the far-field emission profile. This \rev{dipole model} streamlines the complex modeling of near-field scattering by leveraging a dipole approximation \cite{zhu2013theoretical}. Paired with a single control simulation using FDTD analysis, this approach enables the precise and efficient tuning of key design parameters. Beyond optimizing our current system, this method holds potential for applications across diverse quantum photonic designs.
\rev{Throughout, the scattering response is modeled at first order (electric-dipole approximation) and calibrated to one FDTD control simulation; incorporating higher-order multipole terms is an important direction for future work.}
\rev{Optimization targeted collection efficiency and Gaussian far-field overlap over bounded parameter ranges (Table~\ref{table2}) using Bayesian optimization; robustness was verified via independent parameter sweeps (Figs.~\ref{figure4}--\ref{figure6}).}

\section{Design and analysis}

\begin{figure*}
    \centering
     \includegraphics[width=\textwidth]{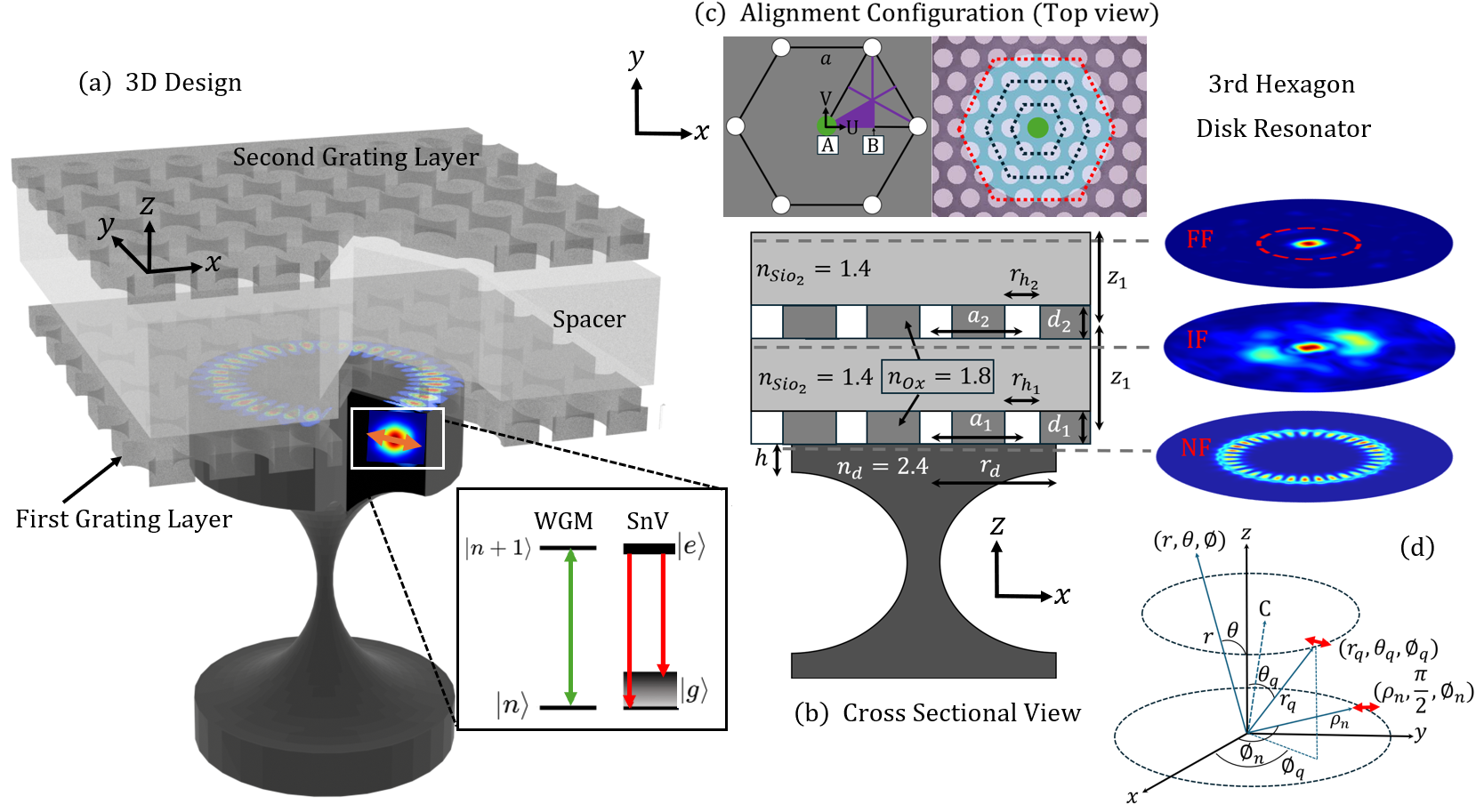}
     \caption{(a) 3D Design: A schematic representation of a spin-photon interface consisting of a disk resonator coupled to a dual-grating structure. The orange mark represents the point dipole
source used to excite the disk mode. (b) Cross-sectional view of the design. On the right, the nearfield (NF), intermediate field (IF), and farfield (FF) are shown at different heights. (c) Hexagonal lattice unit cell and local $(u,v)$ coordinate system. The blue symmetry lines identify a reduced-symmetry domain (purple) used to parameterize alignment; points A and B are key alignment points $(u,v)$ exhibiting high radial symmetry. A zoom-in highlights the WGM–grating interaction region. Expanding hexagonal traces from a center lattice point in a triangular lattice: for $n=3$, 18 lattice holes form the perimeter of the hexagon; 10 holes efficiently interact with the WGM of the resonator. The microdisk is shown in light blue}.
    \label{figure1}   
\end{figure*}

The 3D and cross-sectional view of the proposed design is shown in Figure \ref{figure1} (a) and (b). It consists of a microdisk resonator incorporating an SnV color center, integrated with two hexagonal perturbation grating layers. These layers, composed of silicon oxynitride (\(n_{\text{Ox}} = 1.8\)), are positioned above the microdisk. The first grating layer is placed in the near-field, directly above the microdisk resonator, and the second is in the intermediate-field region. A silicon dioxide (\(n_{\text{SiO}_2}=1.4\)) spacer separates the two layers. \rev{Figure \ref{figure1}(c) annotates the local $(u,v)$ coordinate system and highlights a zoom-in of the WGM–grating interaction region used in our alignment analysis.}
Several geometric parameters can be optimized to achieve both the desired Gaussian far-field profile and high efficiency in performance. These parameters are defined in Table \ref{table1}.

\begin{table}[h]
    \centering
    \begin{tabular}{ll} 
        \toprule
         \\
        \midrule
        $r_d$ &  Radius of Disk Resonator \\
        $h$ & Thickness of Disk Resonator  \\
        $a_1$ & Lattice Constant, First Grating Layer  \\
        $a_2$ & Lattice Constant, Second Grating Layer   \\
        $d_1$ & Hole Height, First Grating Layer  \\
        $d_2$ & Hole Height, Second Grating Layer \\
        $r_{h1}$ & Hole Radius,First Grating Layer  \\
        $r_{h2}$ &  Hole Radius, Second Grating Layer \\

        \bottomrule
    \end{tabular}
     \caption{List of geometric parameters for general structure.}
    \label{table1}
\end{table}

The design is robust to alignment errors between the scatterer's holes and the microdisk. The range of possible misalignment configurations is represented by the purple region in Figure \ref{figure1} (c). When the green reference hole is located at the origin of the local coordinate system (U, V), the system is fully aligned with the optical axis of the microdisk. In the full alignment configuration, the scatterer holes are symmetrically distributed around the optical axis of resonator. Any displacement of the green hole within the purple triangle results in a misalignment.  

We first present an analytical approach for the symmetric case, where the dipoles symmetrically overlap with the WGM of the cavity. Next, we numerically analyze the structure using the \rev{dipole model}. Finally, we compare these results with those obtained from FDTD simulations. 
  
When the hole sizes are much smaller than the resonance wavelength of the resonator, the scattered field at each hole is well-approximated by dipole radiation. However, when the hole sizes become comparable to the resonance wavelength, higher-order multipole contributions must be considered.  
To compute field emission, we model each scattering element (hole) as a dipole with a  dipole moment 
\(
\mathbf{p}_n = \alpha \mathbf{E}_n
\)
where \(\alpha\) is the polarizability determined by the scatterer's geometry, and \(\mathbf{E}_n\) is the local electric field at the scatterer \(n\). The dipoles oscillate with \(\omega\), the angular frequency of the optical field.

We choose the reference coordinate system to be located at the center of the first perturbation layer. The dipoles in the first grating layer are positioned at  spherical coordinate
\( \left(  \rho_n, \pi/2, \phi_n \right)
\) where \(\rho_n\) is the distance of dipole \(n\) from the origin, and \(\phi_n = \frac{2\pi n}{N}\) defines the azimuthal position of the dipoles.  
The dipoles in the second perturbation layer are located at  
\(
\left( r_q, \theta_q, \phi_q\right)
\) (Figure \ref{figure1}(d)). 

In full alignment case, the \(l\)-th hexagon trace interacts with a circle with radius \(l a\), where \(a\) denotes the lattice constant. For each \(l\)-th hexagon, the \(6l\) dipoles are evenly spaced, with a separation distance of \(a\) between adjacent dipoles (Figure \ref{figure1} (c)). 
To calculate the field pattern in the upper hemisphere, each time we consider the collective contributions of \(N=6 l\) dipoles, uniformly distributed along the \(l\)-th hexagon trace. The total field is then obtained by superposing the fields from all hexagon traces, i.e.,



\begin{equation}
\begin{aligned}
\mathbf{E} &= \sum_{n=1}^{N} \frac{1}{4\pi \epsilon_0} \Bigg\{ 
\frac{\omega^2}{c^2 R_n} 
\Big[ ( \hat{\mathbf{R}}_n \times \mathbf{p}_n ) \times \hat{\mathbf{R}}_n \Big] \\
&\quad + \left( \frac{1}{R_n^3} - \frac{i \omega}{c R_n^2} \right) 
\Big[ 3\hat{\mathbf{R}}_n ( \hat{\mathbf{R}}_n \cdot \mathbf{p}_n ) - \mathbf{p}_n \Big] 
\Bigg\} e^{i k R_n}.
\label{equation2}
\end{aligned}
\end{equation}

where \(\mathbf{R}_n = R_n \hat{\mathbf{R}}_n = \mathbf{r} - \mathbf{r}_n\) is the vector from the \(n\)th dipole to the observation point \(\mathbf{r}\). The wave number is given by \(k = \frac{2\pi}{\lambda}\).





 In the near-field (NF) regime, where \( k R_n \ll 1 \), the dominant terms scale as \( (k R_n)^{-3} \). Since the fields are short-range, the primary contribution arises from the interference of adjacent dipoles.  

In the intermediate-field (IF) regime, where \( k R_n \approx 1 \), the dominant terms scale as \( (k R_n)^{-2} \).  In the far-field (FF) regime, where \( k R_n \gg 1 \), only the \( (k R_n)^{-1} \) terms persist. In both the intermediate and far-field regimes, the field pattern is primarily shaped by the interference of radiation emitted by all dipoles collectively.

We first compute the intermediate field by considering the interaction of the near-field with the first perturbing layer. At intermediate distance, we use Fresnel approximation.
To calculate the emission of the \( E_{\rho} \)-polarized mode, the electric dipoles are assumed to be aligned along the \(\rho\)-direction. Similarly, for the \( E_z \)-polarized mode, the dipoles are oriented along the \( z \)-direction.

For \(E_{\rho}\)-polarized mode, the dipole moment of \(n\)-th dipole is  
\(
\mathbf{p}_n = \alpha_1 {E}_{\rho}^{\text{NF}}(\rho_n) \hat{\mathbf{\rho}}_n,
\) 
where \( {E}_{\rho}^{\text{NF}} (\rho_n)\) represents \(\rho\) component of nearfield at the position of the \(n\)th scatterer and \(\alpha_1\) is the polarizability of the scatterers in the first grating layer and assumed to be identical for all scatterers.
Moreover, the unit vector associated with the \(n\)th dipole in paraxial approximation limit is given by  
\(
\hat{\mathbf{R}}_n = \hat{\mathbf{\rho}}_n + \hat{z}.
\) 
 Since both WGM with mode number \(M\) and the interacting hexagon with \(N\) holes are periodic,  the amplitude of nearfield at the place of dipoles is periodic with a difference frequency of \(L= M - N \), and \(L\) is called the topological Pancharatnam
charge.  
Therefore, the electric field at the intermediate distance at a point \(\text{C} (r,\theta, \phi)\) is expressed as:
\begin{equation}
\begin{aligned}
\mathbf{E}^{\text{IF}} &= \frac{-i \omega  \alpha_{1}}{4\pi \epsilon_0 c} 
\sum_{n=1}^{N}  \frac{ e^{i k R_n}}{ R_n^2} e^{iL\phi_n} 
\Big( 2\hat{\rho}_n + 3 \hat{z} \Big) \\
&\quad \times {E}_{\rho}^{\text{NF}} (\rho_n)
\end{aligned}
\label{equation3}
\end{equation}

Then, in the intermediate distance, the \(\rho\) component \(\text{E}^{\text{IF}}_{\rho}= \hat{\rho}.\mathbf{E}^{\text{IF}} \), \(\phi\) component \(\text{E}^{\text{IF}}_{\phi}= \hat{\phi}.\mathbf{E}^{\text{IF}} \) and \(z\) component \(\text{E}^{\text{IF}}_{z}= \hat{z}.\mathbf{E}^{\text{IF}} \) are respectively obtained as follows:

\begin{align}
  \text{E}^{\text{IF}}_{\rho} &= \frac{-2i \omega  \alpha_{1}}{4\pi \epsilon_0 c} 
  \sum_{n=1}^{N}  \frac{ e^{i k R_n}}{ R_n^2} e^{iL\phi_n} \cos(\delta \phi_n) \text{E}_{\rho}^{\text{NF}} (\rho_n) \label{eq:Erho} \\
  \text{E}^{\text{IF}}_{\phi} &= \frac{-2i \omega  \alpha_{1}}{4\pi \epsilon_0 c} 
  \sum_{n=1}^{N}  \frac{ e^{i k R_n}}{ R_n^2} e^{iL\phi_n} \sin(\delta \phi_n) \text{E}_{\rho}^{\text{NF}} (\rho_n) \label{eq:Ephi} \\
  \text{E}^{\text{IF}}_{z} &= \frac{-3i \omega  \alpha_{1}}{4\pi \epsilon_0 c} 
  \sum_{n=1}^{N}  \frac{ e^{i k R_n}}{ R_n^2}e^{iL\phi_n} \text{E}_{\rho}^{\text{NF}} (\rho_n) \label{eq:Ez}
\end{align}

where \(\delta \phi_n=\phi_n-\phi\). Under the Fresnel approximation and assuming a high density of scatterers such that \(2 \pi/N \ll1\), we can approximate the summation using an integral representation: \(\sum_{n=1}^{N} f(\delta \phi_n)\approx \frac{N}{2\pi}\int_{0}^{2\pi} f(\phi) d\phi\). Moreover, we assume that the amplitude of the electric field is constant for all dipoles. Based on these assumptions, the field components at a point \((r,\theta, \phi)\) where \(r=\sqrt{\rho^2+z^2}\) are simplified as follows:
\begin{flalign}
  \label{equation7}
  \text{E}^{\text{IF}}_{\rho} &=  \frac{-2(i)^L \text{f}(\rho, z)\text{N}L}{k\rho_n \tan\theta} 
  e^{iL\phi} J_{L}(k\rho_n \tan\theta) &\\
  \text{E}^{\text{IF}}_{\phi} &= 2(i)^{L-1}\text{f}(\rho, z)\text{N} e^{iL\phi} J'_{L}(k\rho_n \tan\theta) &\\
  \text{E}^{\text{IF}}_{z} &= 3(i)^{L}\text{f}(\rho, z)\text{N} e^{iL\phi} J_{L}(k\rho_n \tan\theta) &
\end{flalign}
where 

\begin{align}
    \text{f}(\rho, z)=\frac{-i \omega  \alpha_{1}\text{E}_{\rho}^{\text{NF}} (\rho_n) }{4\pi \epsilon_0 c} \frac{ e^{i k z(1+\frac{\rho^2}{2z^2})}}{z^2}.
\end{align}

By projecting the radial and azimuthal fields into \(x\) and \(y\) directions, the total transverse field is obtained as follows:
\begin{equation}
\begin{aligned}
  &\mathbf{E}^{\text{IF}}_{\text{T}} = 
  \begin{bmatrix}
    E_x^{\text{IF}} \\
    E_y^{\text{IF}}
  \end{bmatrix} 
  = \text{N} \, \text{f}(\rho, z) \Bigg[ \\
  &J_{L-1}(k\rho_n \tan\theta) e^{i(L-1)\phi}  
  \begin{bmatrix}
    1 \\
    i
  \end{bmatrix} \\
  &+ J_{L+1}(k\rho_n \tan\theta) e^{i(L+1)\phi}  
  \begin{bmatrix}
    1 \\
    -i
  \end{bmatrix}
  \Bigg]
\end{aligned}
\end{equation}

This shows that the intermediate field can be described as the superposition of two orthogonal scalar waves: a left-hand circularly polarized (LHCP) beam with a topological charge of \( l-1 \) and a right-hand circularly polarized (RHCP) beam with a topological charge of \( l+1 \) . 
The intermediate field is then perturbed by the second grating layer. To account for this, we introduce a new set of dipole scatterers and compute the resulting field pattern in the far-field region. In farfield regime, we use the Fraunhofer approximation.  In a low numerical aperture (NA) region, the unit vector for all dipoles pointing toward a far-field point can be approximated as \( \hat{\mathbf{d}}_q = \hat{z} \). 

Under this assumption, the \( z \)-component of the intermediate field cannot propagate to the far field. This follows from Equation \ref{equation2},  which states that
\(
(\hat{z} \times \hat{z}) \times \hat{z} = 0.
\)
For the \( \rho \)-component, the dipole moment of the \( q \)th scatterer is given by
\(
\mathbf{p}_q = \alpha_2 {E}_{\rho}^{\text{IF}}(r_q) \hat{\mathbf{\rho}}_q,
\)
where \( {E}_{\rho}^{\text{IF}}(r_q) \) represents the \( \rho \)-component of the intermediate field at the position of the \( q \)th scatterer, and \( \alpha_2 \) denotes the polarizability of the scatterers in the second grating layer.

The interaction of the intermediate field with the second grating layer results in the following far-field expression at a point \( (r, \theta, \phi) \):

\begin{equation}
\begin{aligned}
\mathbf{E}^{\text{FF}} &= \frac{ \alpha_2 \omega^2 }{4\pi \epsilon_0 c^2} 
\sum_{q=1}^{Q} \frac{e^{i k d_q}}{d_q} \Big( \text{E}_{x}^{\text{IF}}(r_q) \hat{x} \\
&\quad + \text{E}_{y}^{\text{IF}}(r_q) \hat{y} \Big)
\end{aligned}
\label{equation11}
\end{equation}

Here, \(d_q\) represents the distance between intermediate dipoles and the far-field region, and \(Q\) refers to the number of dipoles on a hexagon trace within the second grating layer. 
By approximating the phase factor we get:

\begin{equation}
\begin{aligned}
\mathbf{E}^{\text{FF}} &= \frac{ \alpha_2 \omega^2 }{4\pi \epsilon_0 c^2} \frac{e^{i k(r-r_q \cos(\theta)\cos(\theta_q))}}{r} \\
&\quad \times \sum_{q=1}^{Q} e^{-i k r_q\sin(\theta)\sin(\theta_q)\cos(\delta\phi_q)} \\
&\quad \times \Big( \text{E}_{x}^{\text{IF}}(r_q)\hat{x} + \text{E}_{y}^{\text{IF}}(r_q)\hat{y} \Big)
\end{aligned}
\label{equation}
\end{equation}

where we have assumed that \(r_q\) and \(\theta_q\) are equals for all dipoles since all of them are distributed along a circle. Again we approximate the summation by integral.  After simplifications the  total field in far-field are obtained as follows:
\begin{equation}
\begin{aligned}
  &\mathbf{E}^{\text{FF}} = 
  \begin{bmatrix}
    E_x^{\text{FF}} \\
    E_y^{\text{FF}}
  \end{bmatrix} 
  = \text{Q} \, \text{g}(r,\theta) \Bigg[ \\
  &J_{L-1}(k\rho_n \sin\theta_q) J_{L-1}(kr_q \sin\theta \sin\theta_q) e^{i(L-1)\phi} 
  \begin{bmatrix}
    1 \\
    i
  \end{bmatrix} \\
  &+ J_{L+1}(k\rho_n \sin\theta) J_{L+1}(kr_q \sin\theta \sin\theta_q) e^{i(L+1)\phi} 
  \begin{bmatrix}
    1 \\
    -i
  \end{bmatrix}
  \Bigg]
\end{aligned}
\label{equation13}
\end{equation}

where 

\begin{align}
\text{g}(r,\theta)=\frac{ \alpha_2 \omega^2 \text{N}}{4\pi \epsilon_0 c^2} \text{f}(\rho_q,z_q) \frac{e^{i k(r-r_q \cos(\theta)\cos(\theta_q))}}{r}.
\end{align}

The far-field is a superposition of a LHCP wave with topological charge \( L-1 \) and a RHCP wave with topological charge \( L+1 \). The intensity in the far-field is nonzero near the optical axis only when the selection rule is \( L = 1 \), indicating that the overall field pattern is primarily determined by the first grating layer. However, the second grating layer plays a crucial role in suppressing side lobes and concentrating the intensity near the center. 

\begin{figure}
    \centering
     \includegraphics[width=.47\textwidth]{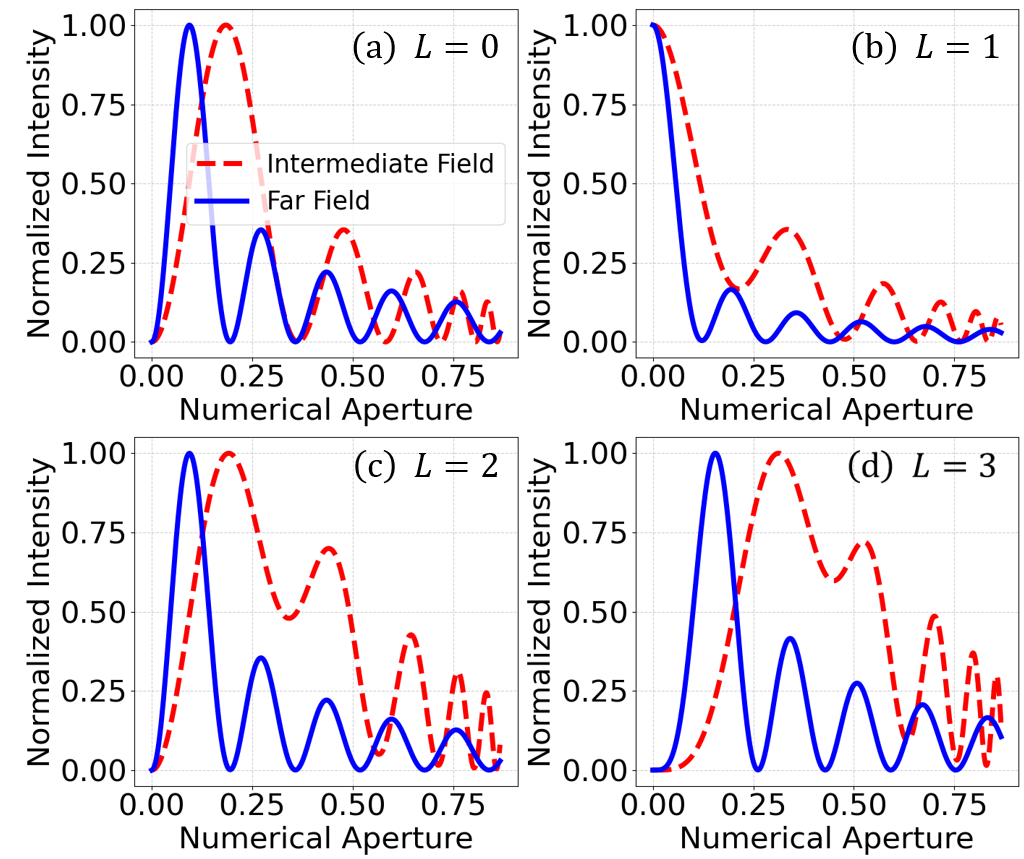}
     \caption{ Cross sections of the intermediate field (dashed red) and farfield (blue) intensity distributions as a function of numerical aperture, with charges:  
(a) \( L = 0 \), (b) \( L = 1 \), (c) \( L = 2 \), and (d) \( L = 3 \).}
    \label{figure2}
\end{figure}

Figure \ref{figure2} presents the analytical model of the normalized intensities for both the intermediate and far-field distributions across different topological charges. As shown, the second perturbation layer by destructive interference minimized light intensity at high NA and effectively concentrates most of the light intensity near the center, even for higher OAM states. This is achieved through an additional modulation factor introduced by the second grating layer, represented by the Bessel function  
\( J_{L\pm1}(k r_q \sin\theta \sin\theta_q) \), which further confines the intensity toward the optical axis.  

This filtering mechanism reduces energy spread and mitigates the central dip observed in single-layer gratings for higher-order OAM states. Furthermore, Equation \ref{equation13} indicates that the far-field distribution remains largely independent of the specific perturbation pattern of the second grating layer. This implies that a similar outcome can be achieved as long as the scatterers are densely distributed in high-intensity regions. These observations are further validated using the \rev{dipole model}.

In a similar way, we can find the emission of the \( E_z \) component of the near field. The emission of this component at the intermediate field is obtained by:

\begin{equation}
\begin{split}
\mathbf{E}^{\text{IF}} = \frac{-i \omega  \alpha_{1}}{4\pi \epsilon_0 c} \sum_{n=1}^{N}  \frac{ e^{i k R_n}}{ R_n^2} e^{L\phi_n} \left( 3\hat{\rho}_n + 2 \hat{z} \right){E}_{z}^{\text{NF}} (\rho_n)
\\  
\label{equation14}
\end{split}
\end{equation}

Where \( {E}_{z}^{\text{NF}} (\rho_n) \) represents the \( z \)-component of the electric field at the location of the \( n \)-th scatterer in the first grating layer. The phase factor and the polarization of this field are similar to that of Equation \ref{equation3} but with different coefficients. Therefore, we can follow the same calculations as before and obtain similar results.
It follows that if the color centers are oriented out of plane, their information can still be transferred to the far field. In contrast, without the intermediate perturbation layer, this transfer would not be possible. Therefore, our approach eliminates the need for precise control over the orientation of the color centers inside the cavity.

After, calculating far-field pattern, the collection efficiency, \( \eta_{col} \), can be assessed for collection into a lens positioned directly above the cavity within a far-field divergence angle of \( \theta_A = \sin^{-1}(\text{NA}) \): 

\begin{equation}
\eta_{col} = \eta_{ex} \frac{\int_0^{\theta_A} \int_0^{2\pi} |\mathbf{E}^{\text{FF}}|^2 \sin(\theta) \, d\phi \, d\theta}{\int_0^{\pi} \int_0^{2\pi} |\mathbf{E}^{\text{FF}}|^2 \sin(\theta) \, d\phi \, d\theta} 
\label{equation15}
\end{equation}

\( \eta_{ex}\) is the extraction efficiency which is the proportion of light extracted from the cavity into the upper plane compared to emission from the cavity in all directions.

\begin{figure*}
    \centering
     \includegraphics[width=\textwidth]{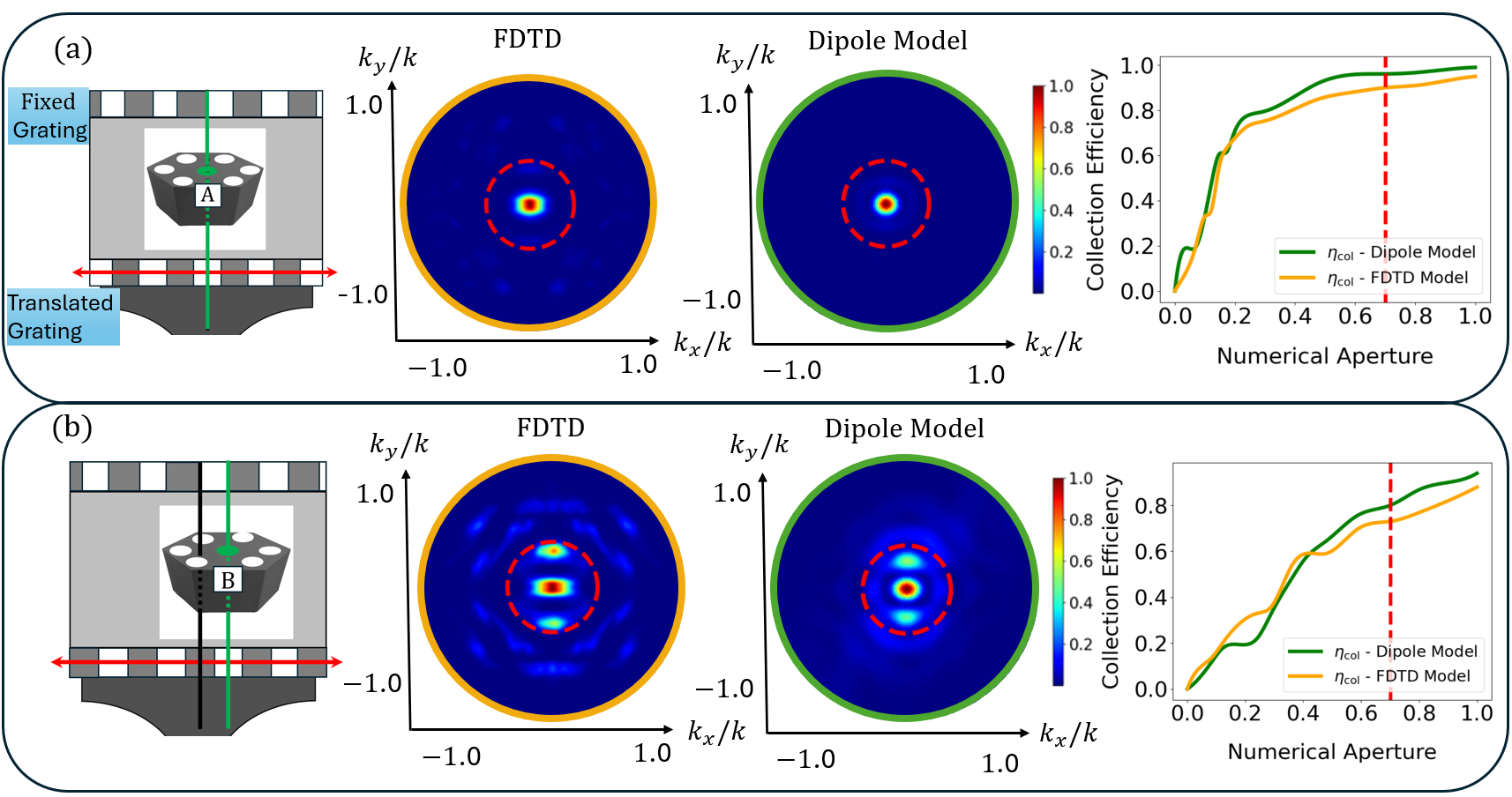}
     \caption{ (a) and (b) illustrate the intermediate and far-field distributions for two different grating alignment configurations. 
    The leftmost schematics depict the structural alignment, where the grating is either fixed or translated relative to the disk resonator. 
    The middle panels show the simulated far-field distributions obtained using the FDTD method and the Dipole Model.
    The rightmost plots present the collection efficiency as a function of the numerical aperture for both models.
    }
    \label{figure3}   
\end{figure*} 

Figure \ref{figure3} shows the field patterns and collection efficiencies when the first grating layer is translated while the second grating layer remains fixed. We examine two specific alignments: full alignment, where the reference scatterer (green hole) is precisely positioned on the optical axis of the disk resonator (configuration \(\text{A}\)); and misalignment, where the reference scatterer is displaced from the optical axis by half of the lattice constant, \(a/2\) (configuration \(\text{B}\)).
 We first optimize the structure for the case, where both the first and second grating layers are fully aligned. The optimized values for both layers are presented in Table \ref{table2}. 

\begin{table}[h]
    \centering
    \begin{tabular}{ll} 
        \toprule
         \\
        \midrule
        Parameters &  Optimized Values \\
        $r_d$ &  \(1.4687 \lambda_0\) \\
        $h$ & \(0.9491 \lambda_0\)  \\
        $a_1$ & \(0.5234 \lambda_0\)  \\
        $a_2$ & \(0.3406 \lambda_0\)    \\
        $d_1$ & \(0.3561 \lambda_0\)  \\
        $d_2$ & \(0.3561 \lambda_0\) \\
        $r_{h1}$ & \(0.1875 \lambda_0\)  \\
        $r_{h2}$ &  \(0.1562 \lambda_0\) \\

        \bottomrule
    \end{tabular}
     \caption{Optimized Values.}
    \label{table2}
\end{table}

We use these optimized values for all other alignment configurations.  
Here, we consider a WGM mode with a mode number of \( M = 17 \), obtained from FDTD simulations. The near-field data is recorded just beneath the first grating layer, as shown in Figure \ref{figure1} (b). For each alignment configuration, we use the corresponding near-field data, since the grating on top of the resonator introduces perturbations, leading to alignment-dependent variations. Different alignment configurations result in varying Purcell factors. For the full alignment case, the Purcell factor of the WGM is 62, while for the misalignment case, it is 66. This leads to an average spectral efficiency of \(\eta_\text{ZPL}=99.6\)\%. 
It should be mentioned that the resonance shift between these configurations is very small, and we neglect this effect in our calculations.

At far-field, the field pattern of full alignment configuration is a Gaussian-like profile. A nearly perfect Gaussian field profile with 95\% overlap is achieved, along with a 96\% collection efficiency within \(\text{NA} = 0.7\). This leads to a total efficiency of \rev{\(\eta_\text{tot}=95.5\)\%} for our definition \(\eta_\text{tot}=\eta_\text{ZPL}\,\eta_\text{col}\). \rev{For reference, the coupling into an ideal Gaussian mode is \(\eta_\text{Gauss}\,\eta_\text{col} \approx 0.95\times 0.96 \approx 91\%\); including spectral weighting gives \(\eta_\text{ZPL}\,\eta_\text{Gauss}\,\eta_\text{col} \approx 90.7\%\).} On the other hand, for misalignment configuration, we still have high concentric intensity near the center and 82\% collection efficiency is achieved which leads to a total efficiency of \(\eta_\text{tot}=81.6\)\%.

\rev{The optimized radii of the first and second scattering holes are \(120~\mathrm{nm}\) and \(100~\mathrm{nm}\), corresponding to \(\lambda_0/6\) and \(\lambda_0/7\) at \(\lambda_0 = 640~\mathrm{nm}\). As shown in the comparative analysis in Figure~3, the dipole model efficiently predicts the far-field pattern. However, to further improve the model’s accuracy, the inclusion of higher-order multipole contributions should be considered.}

\begin{figure*}
    \centering
     \includegraphics[width=\textwidth]{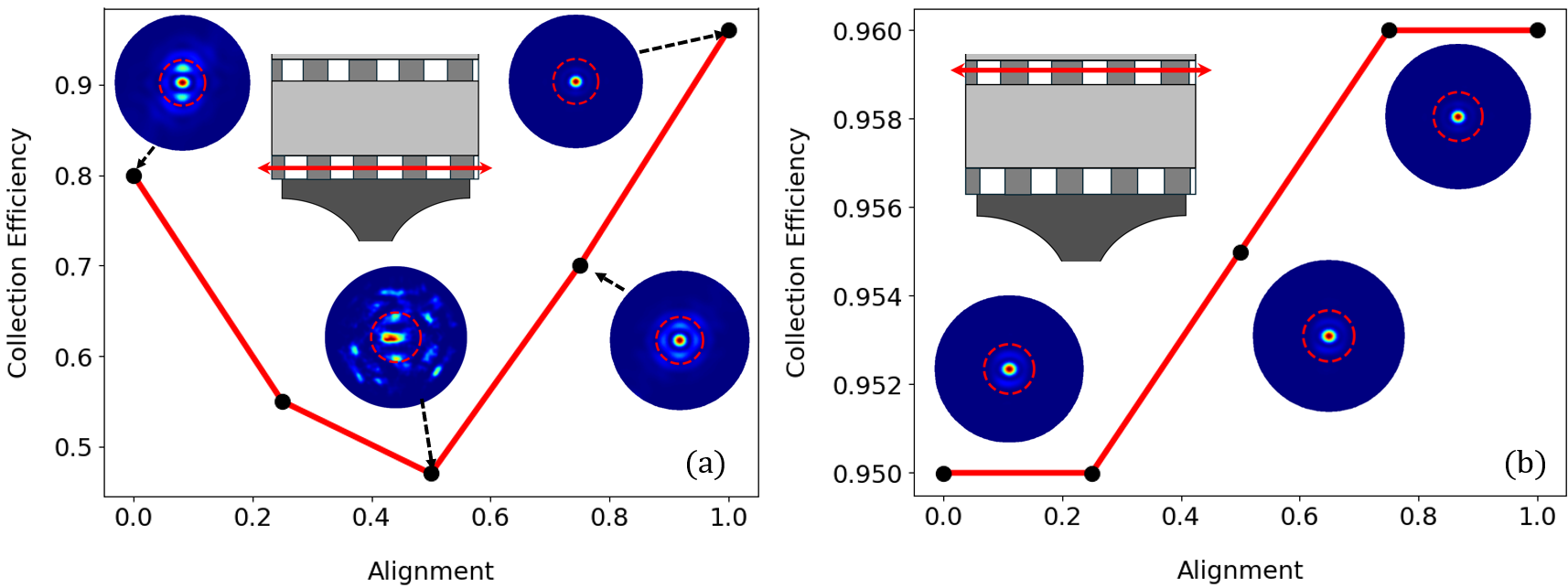}
     \caption{Collection efficiency as a function of the grating alignment.  
     (a) The collection efficiency for different alignments of the first grating layer, while the second grating layer is fixed. The farfield pattern is shown for each alignment configuration.  
     (b) The collection efficiency for different alignments of the first grating layer, while the second grating layer is fixed. The farfield pattern is shown for each alignment configuration. \rev{We parameterize alignment by $s=\Delta/a \in [0,1]$, where $\Delta$ is the lateral offset of the reference scatterer from the optical axis and $a$ is the lattice constant. $s=0$ is perfect alignment (Configuration A in Fig.~\ref{figure1}(c)), and $s=1$ is full misalignment (Configuration B in Fig.~\ref{figure1}(c)). For completeness, $\eta_{\mathrm{tot}}$ values differ by $<0.5\%$ from $\eta_{\mathrm{col}}$ over the sweep; numerical values are provided in the text/SI.}}
    \label{figure4}   
\end{figure*} 

We now evaluate the suitability of our design as a spin-photon interface for other color centers, including NV and SiV. Specifically, we estimate the range of acceptable Purcell factors that maintain a total efficiency above 90\%. We consider the fully aligned configuration, where the collection efficiency is \( \eta_\text{col} = 96\%\) and thus the zero-phonon line (ZPL) efficiency is \( \eta_\text{ZPL} = 92\%\).  

For the SnV center, where  
\(
\frac{\Gamma_{\text{tot,0}}}{\Gamma_{\text{ZPL,0}}} - 1 = 0.25,
\)  
Equation~\ref{equation1} gives the required Purcell factor as \( F_p = 4 \). For the SiV center, with  
\(
\frac{\Gamma_{\text{tot,0}}}{\Gamma_{\text{ZPL,0}}} - 1 = 0.66,
\)  
the corresponding \( F_p \) is 9. Finally, for the NV center, where  
\(
\frac{\Gamma_{\text{tot,0}}}{\Gamma_{\text{ZPL,0}}} - 1 = 32.3,
\)  
the required Purcell factor increases to \( F_p = 412 \).

For the SnV and SiV centers, even a low Purcell factor is sufficient to achieve a total efficiency above 90\%.

To evaluate the robustness of the design against manufacturing variations, we computed the field patterns and collection efficiencies for various grating alignments, as illustrated in Figure \ref{figure4}. The analysis on the left corresponds to the case where the first grating layer is translated while the second grating remains fixed. Across most alignments, the collection efficiency exceeds 50\%. In all cases, the intensity remains highly concentrated at the center.  
In the right analysis, the first grating layer remains fixed while the second grating layer is translated. The collection efficiency and far-field pattern exhibit minimal sensitivity to misalignments of the second grating, indicating the design's robustness to such variations. This aligns with analytical predictions. Due to the high density of scatterer holes, a shift in the position of one hole caused by misalignment is effectively compensated by another, preserving the overall optical performance. 

\begin{figure*}
    \centering
     \includegraphics[width=\textwidth]{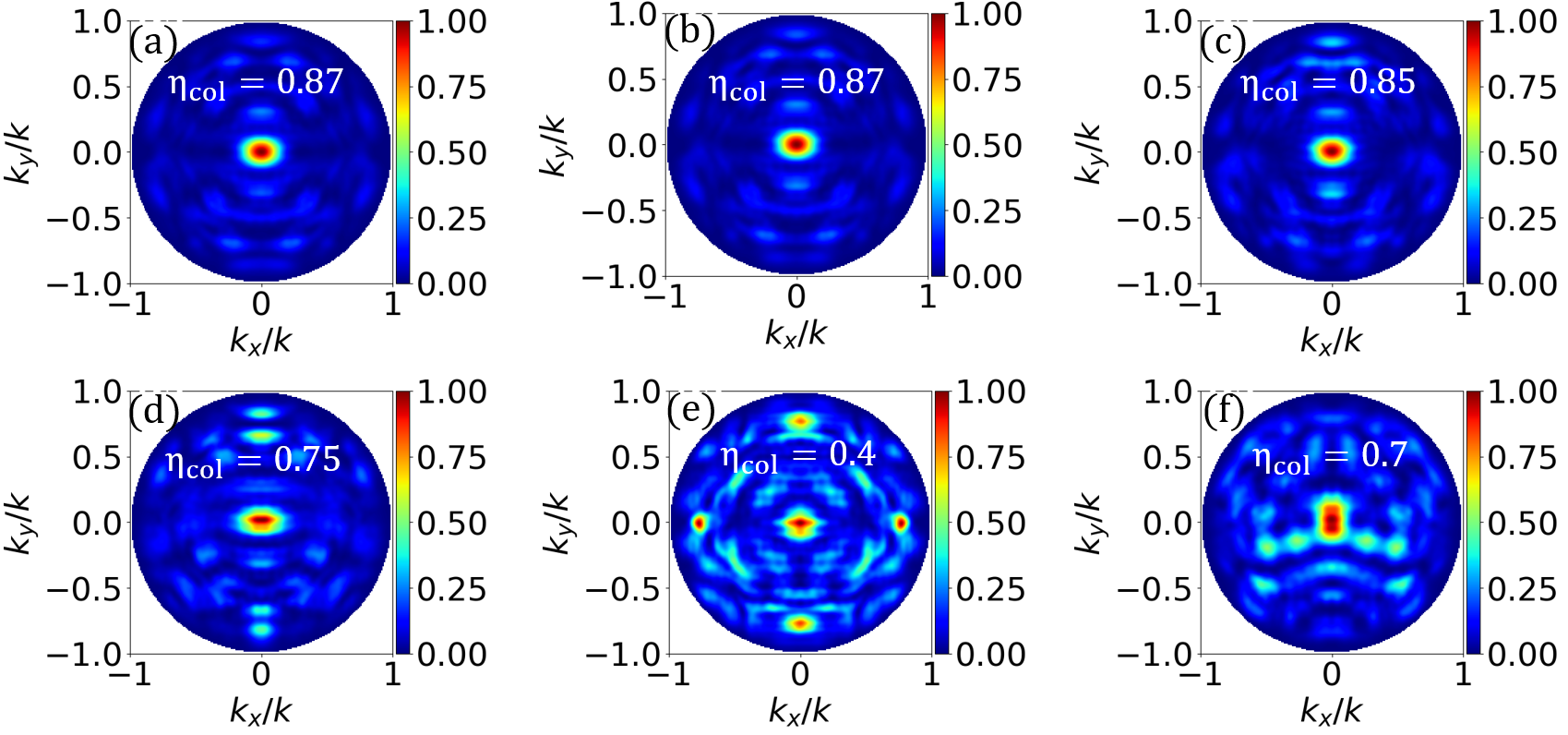}
     \caption{\rev{The far-field distribution for different dipole orientations. (a) \(\theta=90, \phi=90\), (b) \(\theta=45, \phi=90\), (c) \(\theta=25, \phi=90\), (d) \(\theta=10, \phi=90\), (e) \(\theta=0, \phi=90\), (f) \(\theta=90, \phi=0\). }}
    \label{figure5}   
\end{figure*}

\rev{
In this section, we analyze fabrication-related concerns, particularly the effects of non-deterministic dipole orientation of color centers, and variations in the shape, size, and height of scattering holes. 
In practical implementations, the dipole orientation of a quantum emitter cannot be precisely controlled. To address this, one possible approach is to place multiple emitters inside the cavity and selectively excite only those that are aligned with the desired cavity mode. Nevertheless, in our analysis, we also consider the emission profiles for various dipole orientations to evaluate the robustness of the outcoupling performance.
Figure~\ref{figure5} shows the normalized far-field radiation patterns for various dipole orientations. In subfigures (a)--(d), the azimuthal angle is fixed at $\phi = 90^\circ$, while the polar angle $\theta$ varies from $90^\circ$ to $0^\circ$. In case (a), the dipole is oriented entirely in-plane ($\theta = 90^\circ$) and predominantly couples to the TE-like mode, resulting in a symmetric, Gaussian-like far-field profile centered at the origin. As $\theta$ decreases (cases (b)--(d)), the dipole acquires a vertical component, leading to hybrid coupling into both TE- and TM-like modes. This results in increased asymmetry and redistribution of intensity in the far-field pattern. In case (e), where the dipole is oriented vertically ($\theta = 0^\circ$), the emission couples primarily to the TM-like mode, producing multiple high-intensity lobes and a broader angular distribution.  Case (f) corresponds to an in-plane dipole oriented along the $x$-axis ($\theta = 90^\circ$, $\phi = 0^\circ$). Despite minor distortions in some cases, the far-field profiles remain largely Gaussian-like. Moreover, although the device is optimized for TE-mode coupling, this configuration still achieves a collection efficiency of higher than 40\% at an NA of 0.7 for any arbitrary dipole orientation.}

\begin{figure*}[ht]
    \centering
    \includegraphics[width=\linewidth]{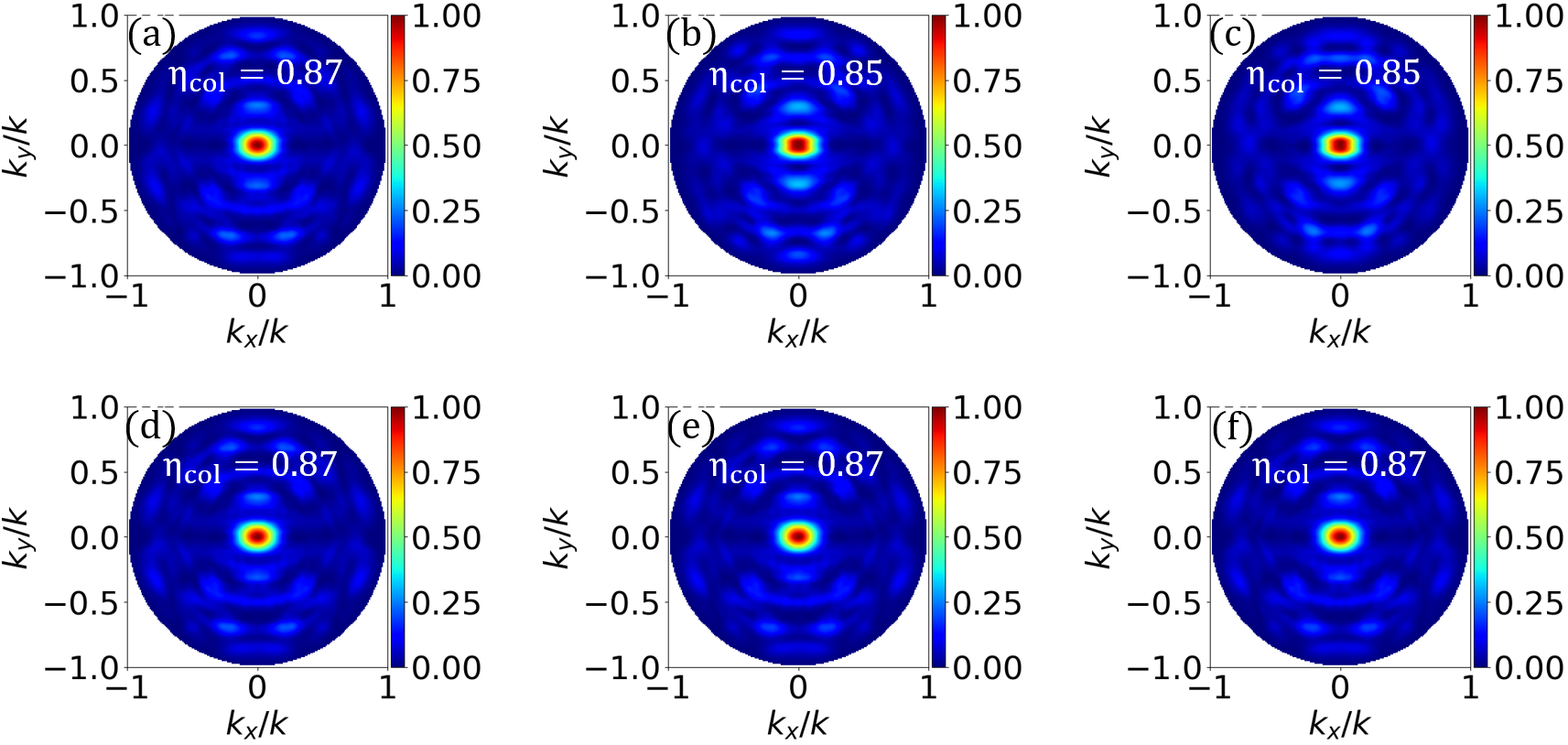}
    \caption{\rev{(a)-(c) The far-field radiation patterns for different scattering hole radii, with identical hole size in both grating layers. The hole thickness is \(220nm\) in all cases. (a) (\(r_{h1}=r_{h2}=140nm\)), (b) (\(r_{h1}=r_{h2}=150nm\)), (c) (\(r_{h1}=r_{h2}=160nm\)). The emission profile remains largely unchanged across variations. (d)-(f) Far-field radiation patterns for different hole heights, with hole radius fixed at 135\,nm in both grating layers. (a) \(d_1 = d_2 = 220\,\mathrm{nm}\), (b) \(d_1 = d_2 = 250\,\mathrm{nm}\), (c) \(d_1 = d_2 = 300\,\mathrm{nm}\). The far-field profiles remain well-centered and symmetric across all cases.}}
    \label{figure6}
\end{figure*}

\rev{Figure~\ref{figure6}(a)--(c) presents the simulated far-field radiation patterns for different scattering hole radii, with identical hole sizes in both grating layers. The variations in hole radius cause only minor changes in the far-field distribution, indicating that the design is robust against moderate fabrication deviations. In all cases, the collection efficiency exceeds 85\%, while maintaining the Gaussian-like profile.}

\rev{Figure~\ref{figure6}(d)--(f) shows the far-field radiation patterns for varying hole heights, while keeping the hole radius fixed at 140\,nm in both grating layers. The far-field profiles and collection efficiency remain stable under these variations.}

\begin{figure*}
    \centering
     \includegraphics[width=\textwidth]{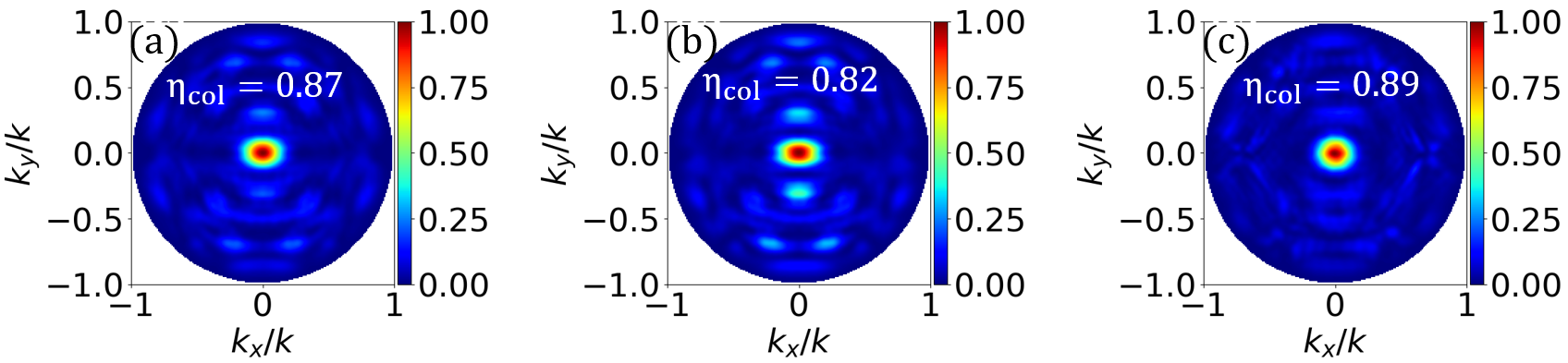}
     \caption{\rev{Far-field radiation patterns for different scattering hole shapes: (a) circular, (b) rectangular, and (c) triangular. All geometries preserve a symmetric and well-centered far-field profile. }}

    \label{figure7}   
\end{figure*}

\rev{Figure~\ref{figure7} illustrates the far-field radiation patterns for different scattering hole geometries: (a) circular, (b) rectangular, and (c) triangular. In all cases, the far-field maintains a symmetric, Gaussian-like distribution.}

\rev{Figure~\ref{figure7} presents the far-field radiation patterns for different scattering hole geometries: (a) circular, (b) square, and (c) triangular. In all cases, the far-field distribution maintains a Gaussian-like profile and maintains high collection efficiency.}

\section{Fabrication considerations}
\label{sec:fabrication}
\rev{Here we outline a representative process flow and practical tolerances for realizing the proposed device, based on established diamond nanophotonics methods and prior demonstrations.}

\textbf{Process flow.} \rev{Diamond microdisks are patterned by electron-beam lithography (EBL) and etched using O\textsubscript{2} ICP-RIE (e.g., \cite{hausmann2014diamond, mouradian2017rectangular, wan2018two}). Early in the flow, Ti/Au alignment marks are defined and reused for all subsequent overlays. The dielectric stack is formed by PECVD deposition of an SiO\textsubscript{2} spacer followed by SiON/SiN layers for the two gratings. Each grating is written with EBL and transferred using CHF\textsubscript{3}/O\textsubscript{2} or CF\textsubscript{4}-based ICP-RIE for SiON/SiN; the diamond layer is unaffected by this step. Closely related diamond bullseye and photonic-crystal processes are reported in \cite{li2015efficient, mouradian2017rectangular, wan2018two}, and a vertically loaded microdisk interface similar to our stack is shown in \cite{duan2021vertically}. Fabrication-tolerant coupling and relaxed alignment strategies in microdisk arrays are discussed in \cite{flores2024alignment}.}

\textbf{Overlay and rotation.} \rev{Modern EBL tools achieve layer-to-layer overlay of \(\lesssim 20\text{--}30\,\mathrm{nm}\) and rotational alignment \(<0.5^{\circ}\) over array fields of a few hundred microns. The second grating in our design is intentionally dense and approximately azimuthally symmetric, so small rotational errors mainly impart a weak azimuthal phase that averages out in the far field, akin to our demonstrated tolerance to lateral translation (Fig.~\ref{figure4}).}

\textbf{Tolerances.} \rev{The additional simulations in Fig.~\ref{figure6} show that variations of \(\pm 10\text{–}20\,\mathrm{nm}\) in hole radius and \(\pm 20\text{–}30\,\mathrm{nm}\) in hole height preserve a well-centered Gaussian-like far field and high collection efficiency. Spacer thickness and inter-layer spacing errors of similar magnitude lead to modest changes, consistent with the design’s robustness to misalignment. The optimized and swept feature sizes (e.g., \(r_{h1}\!\approx\!120\,\mathrm{nm}\), \(r_{h2}\!\approx\!100\,\mathrm{nm}\); 140–160\,nm radius; 220–300\,nm height) are within standard EBL capabilities demonstrated for diamond nanophotonics \cite{li2015efficient, mouradian2017rectangular, wan2018two, burek2014high}.}

\textbf{Materials.} \rev{Silicon oxynitride/nitride on diamond is standard for diffractive outcouplers and photonic-crystal devices; patterning is performed with fluorocarbon chemistries for the dielectric and O\textsubscript{2} ICP for diamond \cite{li2015efficient, wan2018two, mouradian2017rectangular}.}

\rev{Finally, we benchmarked the dipole model against full-wave 3D FDTD simulations. On a university computing cluster (AMD EPYC 7763), the dipole model computes a far-field pattern in 1\,s on a single core, while a full FDTD simulation takes 7\,h using 128 cores—corresponding to a speedup of \(\sim10^6\times\): \(\frac{7\,\mathrm{hr} \times 3600\,\mathrm{s/hr} \times 128\,\mathrm{cores}}{1\,\mathrm{s}} \approx 3.2 \times 10^6\).}

In conclusion, We have presented a high-efficiency vertical emission spin-photon interface based on a dual perturbation grating design integrated with a whispering-gallery mode microdisk resonator. Our approach achieves a 96\% collection efficiency at \(\text{NA} = 0.7\) with a 95\% Gaussian far-field overlap, ensuring compatibility with fiber-based quantum networks.

By introducing two perturbation layers, we enhance far-field control, reduce alignment sensitivity, and improve photon extraction. A \rev{dipole model} validates and accelerates the design process, demonstrating robustness against misalignment and adaptability to different color centers.
This scalable architecture provides a practical solution for quantum communication and quantum memory applications, paving the way for efficient spin-photon interfaces in future quantum networks.

\rev{Finally, we note qualitative limits set by extraction--Q trade-offs and residual higher-order multipole leakage beyond the dipole order; these practical considerations inform achievable efficiencies in fabricated devices.}

\bibliographystyle{quantum}

\end{document}